\documentclass[showpacs,showkeys,preprintnumbers,amsmath,amssymb,twocolumn,prl,floatfix]{revtex4}

\usepackage{amssymb,amsfonts}
\usepackage{amsmath}
\usepackage[dvips]{graphicx}
\usepackage{graphicx}
\begin{document}

\title{Quantum search using non-Hermitian adiabatic evolution }

{\author{Alexander I. Nesterov}
   \email{nesterov@cencar.udg.mx}
\affiliation{Departamento de F{\'\i}sica, CUCEI, Universidad de Guadalajara,
Av. Revoluci\'on 1500, Guadalajara, CP 44420, Jalisco, M\'exico}

\author{Gennady P.  Berman}
 \email{ gpb@lanl.gov}
\affiliation{Theoretical Division, Los Alamos National Laboratory,
Los Alamos, NM 87544, USA }

\date{\today}

\begin{abstract}
We propose a non-Hermitian quantum annealing algorithm which can be useful for solving complex optimization problems. We demonstrate our approach on Grover's problem of finding a marked item inside of unsorted database. We show that the energy gap between the ground and excited states depends on the relaxation parameters, and is not exponentially small. This allows a significant reduction of the searching time. We discuss the relations between the probabilities of finding the ground state and the survival of a quantum computer in a dissipative environment.
\end{abstract}

\pacs{03.67.Ac, 03.67.Lx, 75.10.Nr, 64.70.Tg}

\keywords{critical points; ground states; quantum theory;  adiabatic quantum computation; quantum annealing}
\preprint{LA-UR-12-24262}

\maketitle


Many physical and combinatorial problems associated with complex networks
of interacting degrees of freedom can be mapped to equivalent problems of
finding the minimum of cost function or the ground state of a corresponding quantum
Hamiltonian, $\mathcal H_0$, \cite{FGGLL,KN,SSMO,DC,SMTC,SST,CFS,SNS,AM}.
One of the approaches to find the ground state of  $\mathcal H_0$ is quantum annealing (QA)  which can be formulated as follows. Consider the time-dependent Hamiltonian
$\mathcal H(t) =\mathcal H_0 +  \Gamma(t)\mathcal H_1$,
where $\mathcal H_0$ is the Hamiltonian to be optimized, $\mathcal H_1$ is an auxiliary ``initial" Hamiltonian, and $[\mathcal H_0,\mathcal H_1] \neq 0$.  The coefficient, $\Gamma(t)$, is a control parameter, and $\Gamma(t)$ decreases from very high value to zero during the evolution.

One starts with the ground state of $\mathcal H_1$ as the initial state, and if $\Gamma(t)$ is slowly decreasing, the adiabatic theorem guarantees approaching the ground state of $\mathcal H_0$, at the end of the computation, assuming that there are no energy level crossings between the ground and excited states. So, the quantum optimization algorithms require the presence of a gap between the ground state and first excited state. However, in typical cases the minimal gap, $g_m$, is exponentially small. For instance, in the commonly used quantum optimization $n$-qubit models, the estimation of the minimal energy gap yields: $g_m \approx 2^{-n/2}$ \cite{FGGLL,DC,SUD,JKKM,YKS}. This increases drastically the total computational time, and from a practical point of view the advantage of the method is lost.

Recently \cite{BN}, we have proposed a non-Hermitian adiabatic quantum optimization
with the non-Hermitian auxiliary Hamiltonian. We have shown that the non-Hermitian quantum annealing (NQA) provides an effective level repulsion for the total Hamiltonian. This effect enables us to develop an adiabatic theory without the usual gap condition and to determine the low lying states of $\mathcal H_0$, including the ground state. Some interesting suggestions for implementation of non-Hermitian architectures by realization of ``Ising machine" based on mutually injection-locked laser systems were recently discussed in \cite{Laser1,Laser2}.

In this Letter, we apply the NQA to Grover's problem \cite{GLOK}, i.e. finding a marked item in an unstructured database.

Consider a set of $N= 2^n$ unsorted items among which one item is marked. The related Hilbert space is of dimension $N$. In this space, the basis states are written as $|i\rangle$ (i=1,2,\dots,N), and the marked state is denotes as $|m\rangle$. The task is to find the marked item as rapidly as possible.

The Hamiltonian  whose ground state is to be found, can be written as:
$    {\mathcal H}_0 =    - |m\rangle \langle m|$. Its ground state, marked as $|m\rangle$, is unknown. The auxiliary Hamiltonian is given by
$    {\mathcal H}_1 =- |\psi_0\rangle \langle \psi_0|$,
where $|{\psi}_0\rangle = ({1}/{\sqrt{N}})\sum^N_{i=1} |i\rangle $ is its ground state with energy $E^g_1=-1$. For both Hamiltonians, ${\mathcal H}_0$ and ${\mathcal H}_1$, the rest of eigenstates have the  $N-1$-times degenerate energy $E_r=0$ ($r=2,3, \dots, N$).  (Our choice of the Hamiltonian is different from the Hamiltonian considered in refs. \cite{RN,FGG,FEGS,SG} by a total shift on the unit matrix.) The total time-dependent non-Hermitian Hamiltonian is chosen as follows: ${\mathcal H}_\tau(t) =  {\mathcal H}_0 +  h(t){\mathcal H}_1$,
where
\begin{align}
h(t) = \left\{
\begin{array}{l}
\gamma (\tau-t), \quad 0 \leq t \leq \tau \\
0 , \quad t \geq \tau
\end{array}
\right.
\end{align}
We denote $\gamma=( g+i\delta)/\tau$, where $g$ and $\delta$ are real. In what follows we assume that $\delta \ll 1$.

The adiabatic quantum search algorithm consists of (i) preparing the system in the initial state, $|\psi(0)\rangle =|\psi_0\rangle$, and (ii) performing an evolution by applying the non-Hermitian Hamiltonian, $ {\mathcal H}_\tau(t)$, during a time, $\tau$. At the end of evolution, the non-Hermitian part of the total Hamiltonian disappears. Then, if the evolution is sufficiently slow, the system is remained in its ground state, which will be the ground state of the Hermitian Hamiltonian, ${\mathcal H}_0$.

We start with the solution of the eigenvalue problem for $  {\mathcal H}_\tau(t)$. This yields (N-2)-times degenerate highest eigenvalue, $  E_2 =0$, and two lowest eigenvalues, $  E_0$ and $  E_1$, which are given by
\begin{align}\label{E1b}
&   E_0(t) = -\frac{\varepsilon(t)}{2} - \frac{\Omega(t)}{2}, \\
 &  E_1(t) = -\frac{\varepsilon(t)}{2} + \frac{\Omega(t)}{2},
\end{align}
where $  \Omega(t) = \sqrt{  h^2(t) - 2 h(t) \cos\alpha + 1}$ and $\varepsilon(t) = h(t)+1 $. We set $\sin(\alpha/2) = 1/\sqrt{N}$. The energy gap between the ground state and the first excited state is given by
 $|\Delta   E(t)|=  \big|\sqrt{{  h}^2(t) - 2  h(t) \cos\alpha +1}\big|$.
For $N \gg 1/\delta$ one can show that the minimum of the energy gap is given by $ |\Delta   E|_{\min}  = {\delta}/\sqrt{g^2 + \delta^2}+ {\mathcal O}({1}/{N})$.

In the two-dimensional subspace spanned by the vectors, $|\psi_0 \rangle$ and $|m \rangle$, we choose an orthonormal basis as $|\psi_0 \rangle$ and $ |\psi_1\rangle = (\sin(\alpha/2)|\psi_0\rangle -|m \rangle )/\cos(\alpha/2)$. We complement it to the basis of the $N$-dimensional Hilbert space by adding $(N-2)$ vectors $|\psi_k\rangle$ $(k= 2,\dots,N-1)$, which form the orthonormal basis of the orthogonal $(N-2)$-dimensional Hilbert subspace. Then, an arbitrary state, $|\Psi(t)\rangle$, can be expanded as $|\Psi(t)\rangle = c_0(t) |\psi_0 \rangle + c_1 (t)|\psi_1 \rangle + \sum^{N-1}_{k=2}c_k(t)|\psi_k  \rangle$.

Inserting this expansion into the Shr\"odinger equation,
$  i {\partial }/{\partial t}|\Psi\rangle =   {\mathcal H}_{\tau}|\Psi(t)\rangle$,
we find that the differential equations for the coefficients, $c_0(t)$ and $c_1(t)$, do not involve the coefficients, $c_k(t)$ ($k=2,\dots, N-1$). Then, effectively the $N$-dimensional problem is exactly reduced to the two-dimensional one. So, it is suffices to confine our attention to the two-dimensional subspace.

Choosing the orthonormal basis as $\big\{|\psi_0\rangle = {\scriptsize \left(
                      \begin{array}{c}
                        0 \\
                        1 \\
                      \end{array}
                    \right)}
, |\psi_1\rangle   = {\scriptsize  \left(
                      \begin{array}{c}
                        1 \\
                        0 \\
                      \end{array}
                    \right )} \big \}$, one can write the corresponding effective (non-Hermitian) Hamiltonian as
\begin{equation}\label{H3c}
 {\mathcal H}_{ef}(t)= -\frac{\varepsilon (t)}{2} +\frac{{\boldsymbol { \Omega }}(t)}{2}\cdot \boldsymbol \sigma,
\end{equation}
where $\boldsymbol {\Omega}(t) = (\sin\alpha,0, h(t)- \cos\alpha)$ is a complex vector, and $\boldsymbol \sigma$ denotes the Pauli matrices.

We denote the (right) instantaneous eigenvectors, corresponding to the eigenvalues, $E_a(t)$, as $|u_{a}(t)\rangle$ ($a=0,1$). One can show that $|u_{0}(0)\rangle = |\psi_0\rangle + {\mathcal O}({1}/{N}) $, and $|u_{0}(t) \rangle \rightarrow |\psi_1\rangle + {\mathcal O}({1}/{N}) $, as $t \rightarrow \tau$.

For the two-level system (TLS) governed by the effective non-Hermitian Hamiltonian (\ref{H3c}), the wave-function can be written as $  |\psi(t)\rangle = c_0(t)|\psi_0\rangle + c_1(t)|\psi_1\rangle $. We assume that the evolution of TLS starts at $t_0 = 0$ in the state $|\psi(0)\rangle = |\psi_0\rangle$. This implies the following initial conditions: $c_0(0)=1$ and $c_1(0)=0$.

Writing $c_{a} = U_{a}(t)\exp \big(\frac{i}{2}\int_0^t\varepsilon(t)dt\big)$, and employing the Schr\"odinger equation for the TLS governed by the effective Hamiltonian, ${\mathcal  H}_{ef}(t)$, we obtain the Weber equation for the new functions $U_a(z)$,
\begin{align}\label{Eq3}
  \frac{d^2}{dz^2}U_{0,1}(z) + \big(\pm\frac{1}{2} - \frac{z^2}{4} - i\nu\Big)U_{0,1}(z) =0,
\end{align}
where $z(t)= e^{i\pi/4}(\gamma(\tau -t)-\cos\alpha)/\sqrt{\gamma}$ and $\nu= \sin^2\alpha/4\gamma$.

Solutions of Weber's equation are given by the parabolic cylinder functions, $D_{-i\nu}(\pm z)$,
\begin{align}\label{Eq4}
& U_{0}(z)= (A D_{-i\nu}(z) + B D_{-i\nu}(-z)) ,\\
& U_{1}(z)=  {\sqrt{i\nu }}( B D_{-i\nu-1}(-z) - A D_{-i\nu-1}(z)).
 \end{align}
The constants, $A$ and $B$, being determined from the initial conditions, are found to be: $A= D_{-i\nu-1}(-z_0)\Gamma(1+i\nu)/\sqrt{2\pi\nu}$ and $B= D_{-i\nu-1}(z_0)\Gamma(1+i\nu)/\sqrt{2\pi\nu}$, where we set $z_0 = z(0)$.

It is assumed that the quantum measurement will determine the state of the quantum system at $t > \tau$. We denote the final state of the system as $|\psi_\tau\rangle$. Then, the probability, $P_n$, of finding the system in a given state, $|n\rangle$, can be written as,
\begin{align}\label{Eq9}
 P_n = \frac{|\langle n|\psi_\tau\rangle|^2}{ |\langle \psi_\tau|\psi_\tau\rangle|^2}.
\end{align}
This yields the (intrinsic) probability of transition $|\psi_0\rangle$ $\rightarrow$ $|\psi_1\rangle $ as
\begin{align}\label{Eq2}
 P_\tau(t)= \frac{|c_1(t)|^2}{|c_0(t)|^2 + |c_1(t)|^2}.
\end{align}
Thus, $ P_\tau(\tau)$ is the probability of the system being in the ground state at the end of the evolution.

Using the functions $U_{0,1}(z)$, we write the probability of transition, $P_\tau(\tau)$, as
\begin{align}\label{Eq2a}
 P_\tau(\tau)= \frac{1}{1+ \frac{|U_0(z_\tau)|^2}{|U_1(z_\tau)|^2}},
\end{align}
 where $z_\tau = z(\tau)= -e^{\pi i/4} \cos\alpha/\sqrt{\gamma}$, and for $N\gg 1$ we obtain: $z_\tau \approx -e^{\pi i/4}/\sqrt{\gamma}$.

To estimate the probability of transition, we apply asymptotic formulas for the parabolic functions. This yields
\begin{align}\label{Eq5}
  \frac{U_0(z_\tau)}{U_1(z_\tau)}\approx -\frac{e^{-\pi \nu/2}e^{-z^2_\tau/2}\Gamma(1+i\nu )}{\sqrt{2\pi\nu i}\Big(1 - \frac{e^{-z^2_0/2}}{\sqrt{2\pi}z_0 }\Big)}.
\end{align}

Inserting (\ref{Eq5}) into Eq. (\ref{Eq2}), we obtain the Landau-Zener formula \cite{LL,ZC} for the Hermitian quantum search $(\delta =0)$,
\begin{eqnarray}\label{P}
P_\tau(\tau) =1-e^{-2\pi\nu},
\end{eqnarray}
where $\nu = (\tau/gN)$. We conclude that $P_\tau(\tau) \approx 1$, if $\tau \geq gN $. Thus, to obtain the probability close to  $1$ to remain in the ground state at the end  of evolution, the computational time should be of order $N$. In fact, this result is equivalent to the  well-known result on the complexity of order $N$ provided by quantum adiabatic evolution approach \cite{FGG}, which is the same as in the classical search algorithm.

For the NQA, assuming $N \gg 1$, we obtain the following rough estimate of the computational time: $\tau \geq (g^2/\delta)\ln N$. Thus, the non-Hermitian quantum search has complexity of order $\ln N$, which is much better than the quantum Hermitian (global) adiabatic algorithm. Also, this complexity is certainly better  than one of the adiabatic local search algorithm that has total running time of order $\sqrt{N}$ \cite{RN}. In Fig. \ref{P2a} we present the results of our numerical simulation. For the Hermitian QA ($\delta =0$) the transition probability is: $P_\tau \approx 3\cdot 10^{-8}$; and for the NQA with weak dissipation, $\delta= 0.0025$, the transition probability is: $P_\tau =1$ ($\tau = 1.5 \cdot 10^4 $).

{\em Nonlinear NQA.} -- We define the survival probability of the lossy system as the trace of the density matrix, $P_s(t) = \rm Tr \rho(t)$. Using the asymptotic formulas for the Weber functions, one can show that for $N \gg 1$, the asymptotic behavior of the survival probability is given by:  $ P_s(t) \approx e^{-\delta t}$. (See Fig. \ref{P2a}.) Then, one can see that the conditions to obtain high probabilities for (i) finding the ground state, leading to inequality, $\tau \geq (g^2/\delta)\ln N$,  and (ii) survival of qubits, $\delta t \leq 1$, are not compatible. A compromise can be found by using a local adiabatic evolution approach \cite{RN}.
\begin{figure}[tbp]
\scalebox{0.175}{\includegraphics{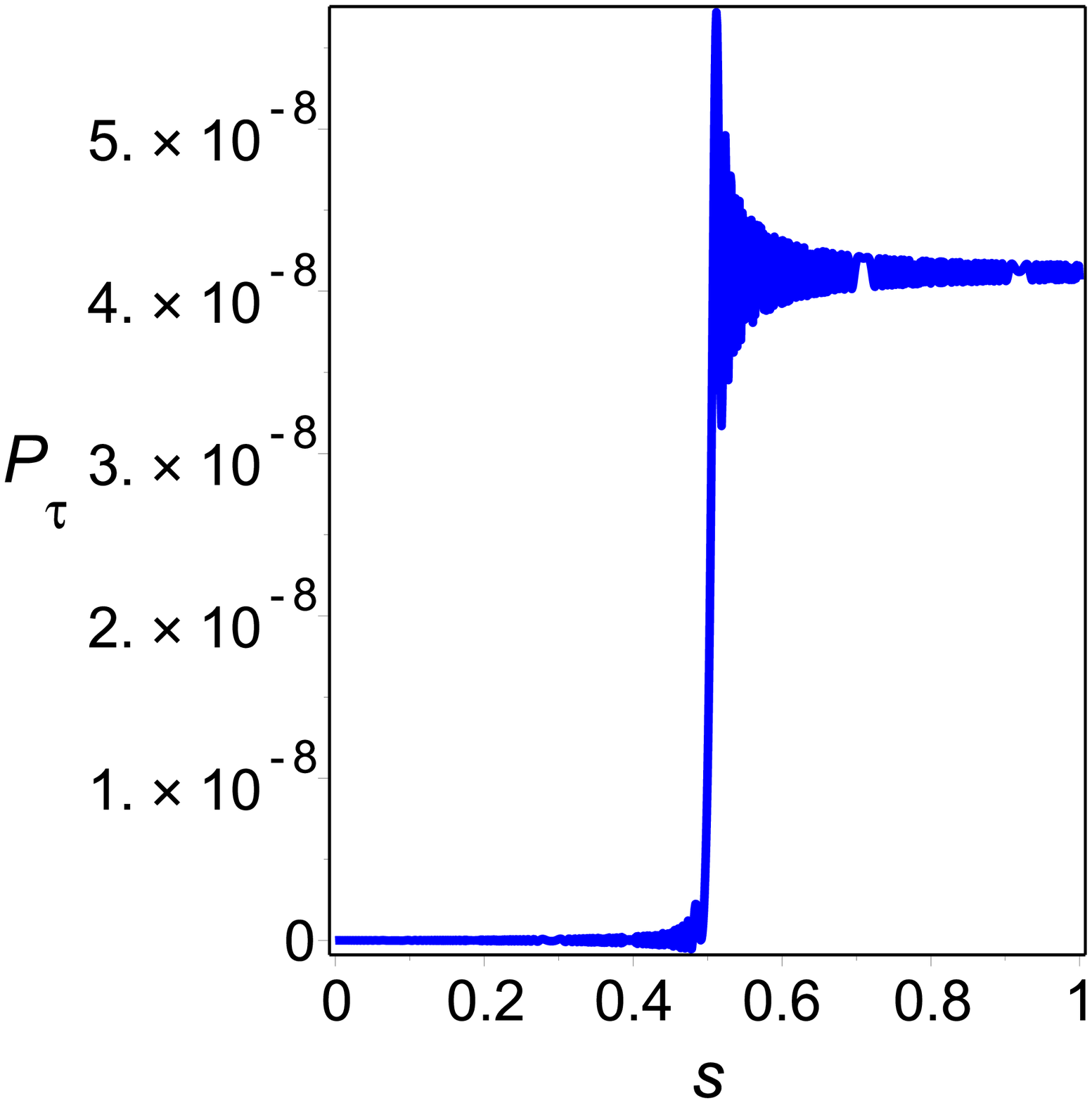}}
\scalebox{0.185}{\includegraphics{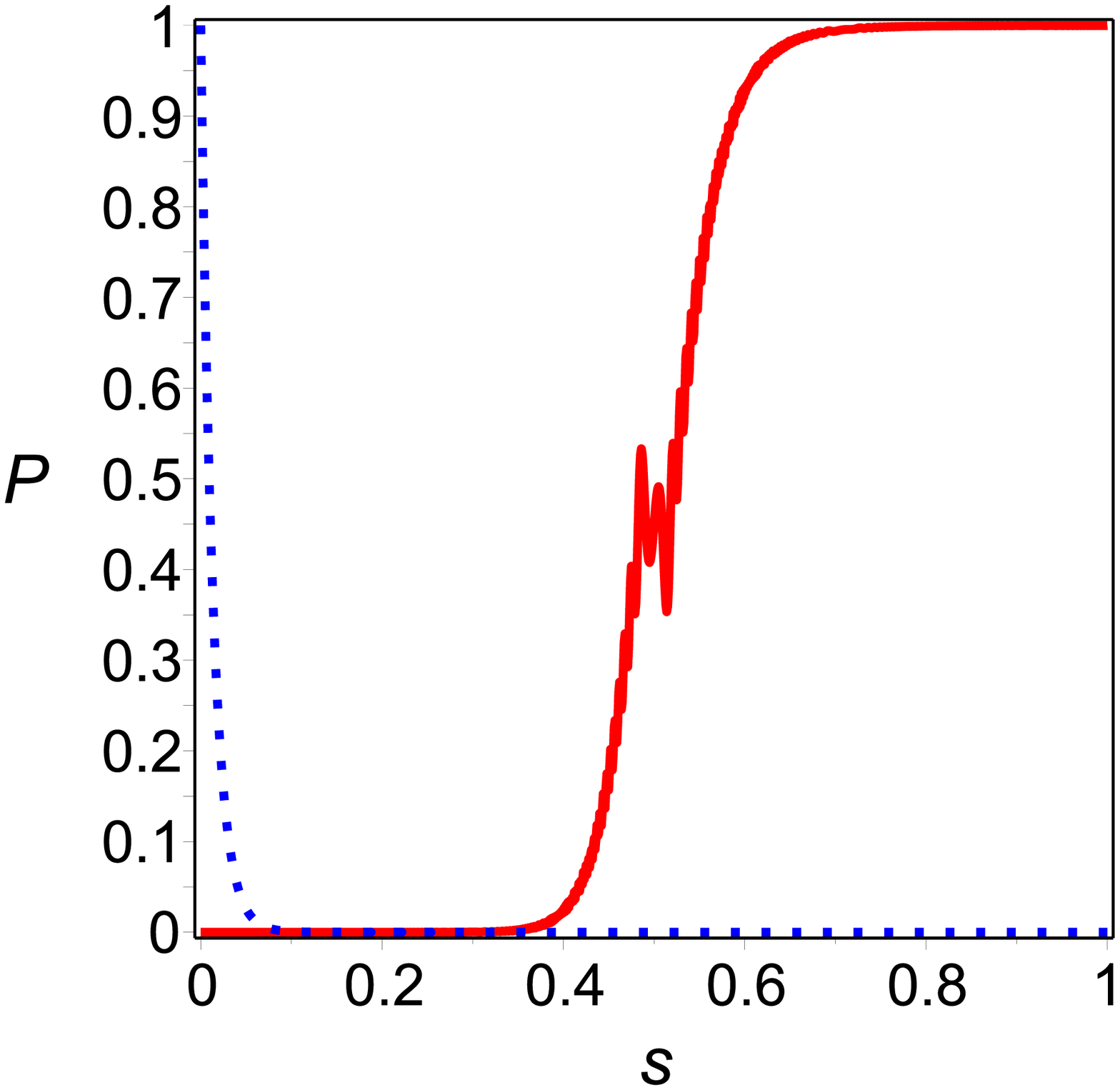}}
\caption{Left panel: The transition probability,  $P_\tau$ as a function of the scaled time, $s =t/\tau $  ($\delta = 0$). Right panel: The survival probability, $P_s(s)$ (dotted blue line), and the transition probability, $P_\tau(s) $ (red line) ($\delta = 0.0025$). In all cases: $g=2,\tau= 1.5\cdot 10^4, N = 2^{40}$.}
\label{P2a}
\end{figure}
\begin{figure}[tbp]
\scalebox{0.18}{\includegraphics{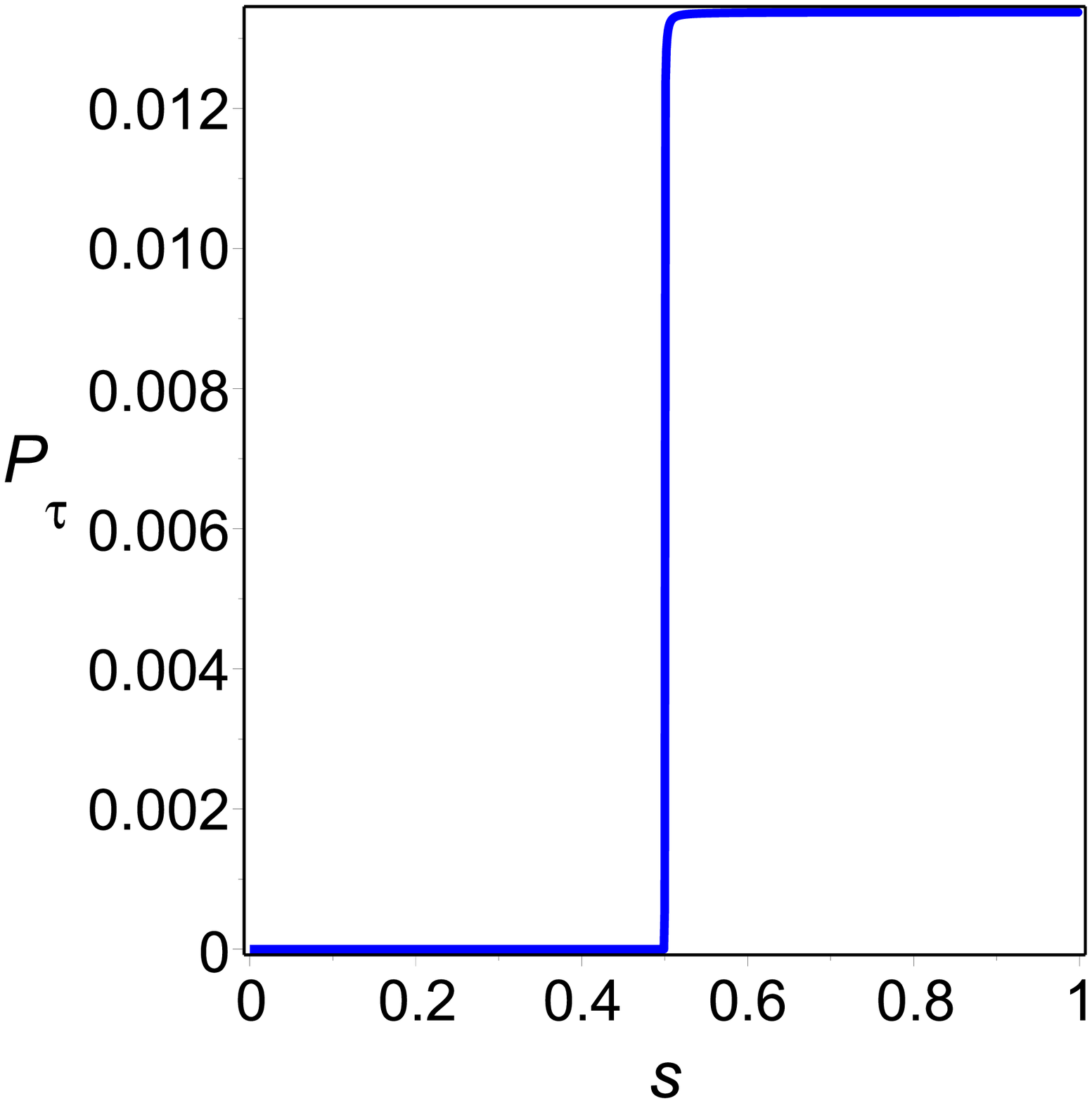}}
\scalebox{0.182}{\includegraphics{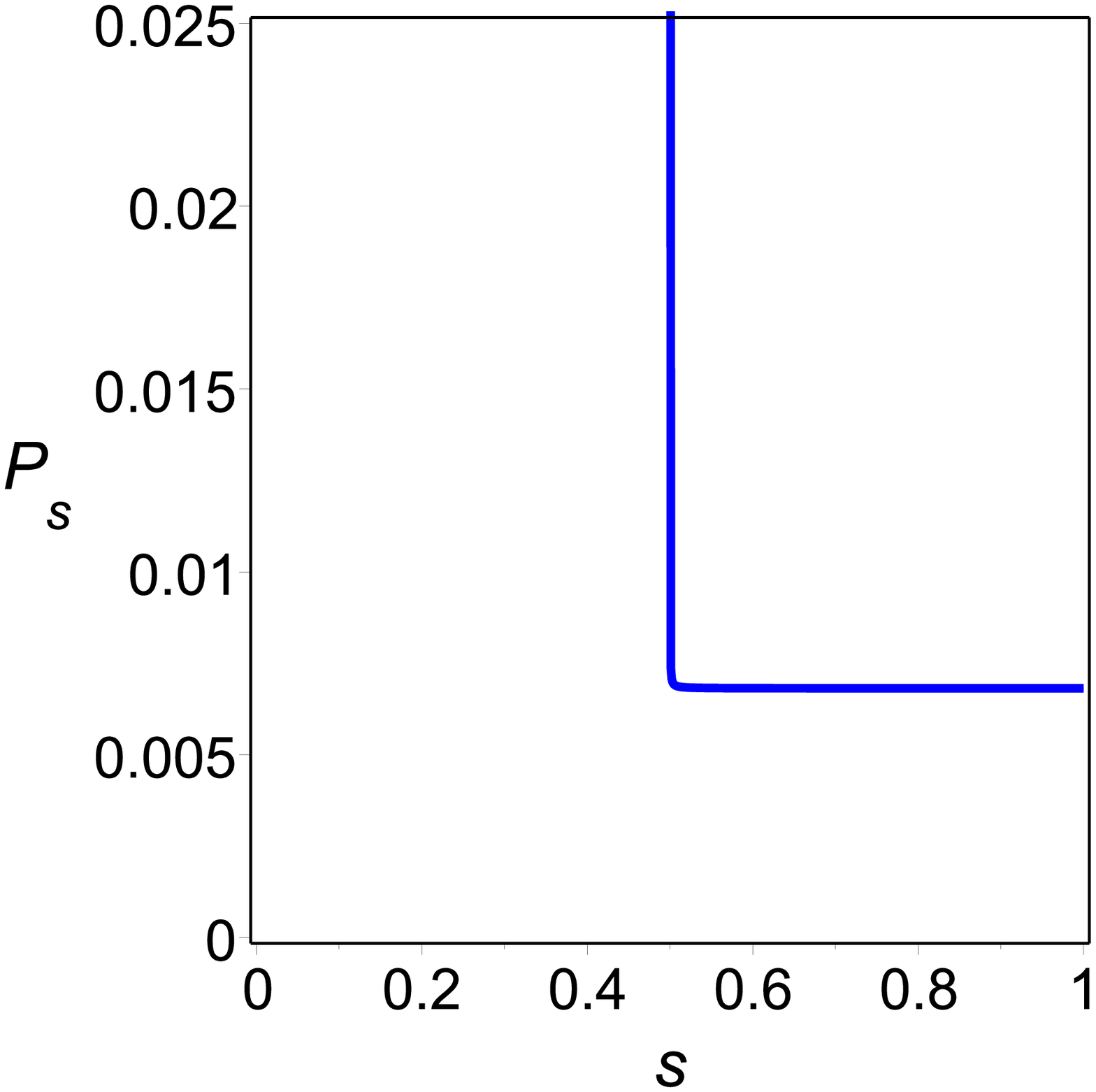}}
\caption{Nonlinear NQA.The transition probability (left panel), and the survival probability (right panel) as the functions of the scaled time, $s =t/\tau$ ($g=2, \delta = 10^{-4}, \tau=  5\cdot 10^{4}, N = 2^{40}$).}
\label{P4}
\end{figure}
\begin{figure}[tbp]
\scalebox{0.18}{\includegraphics{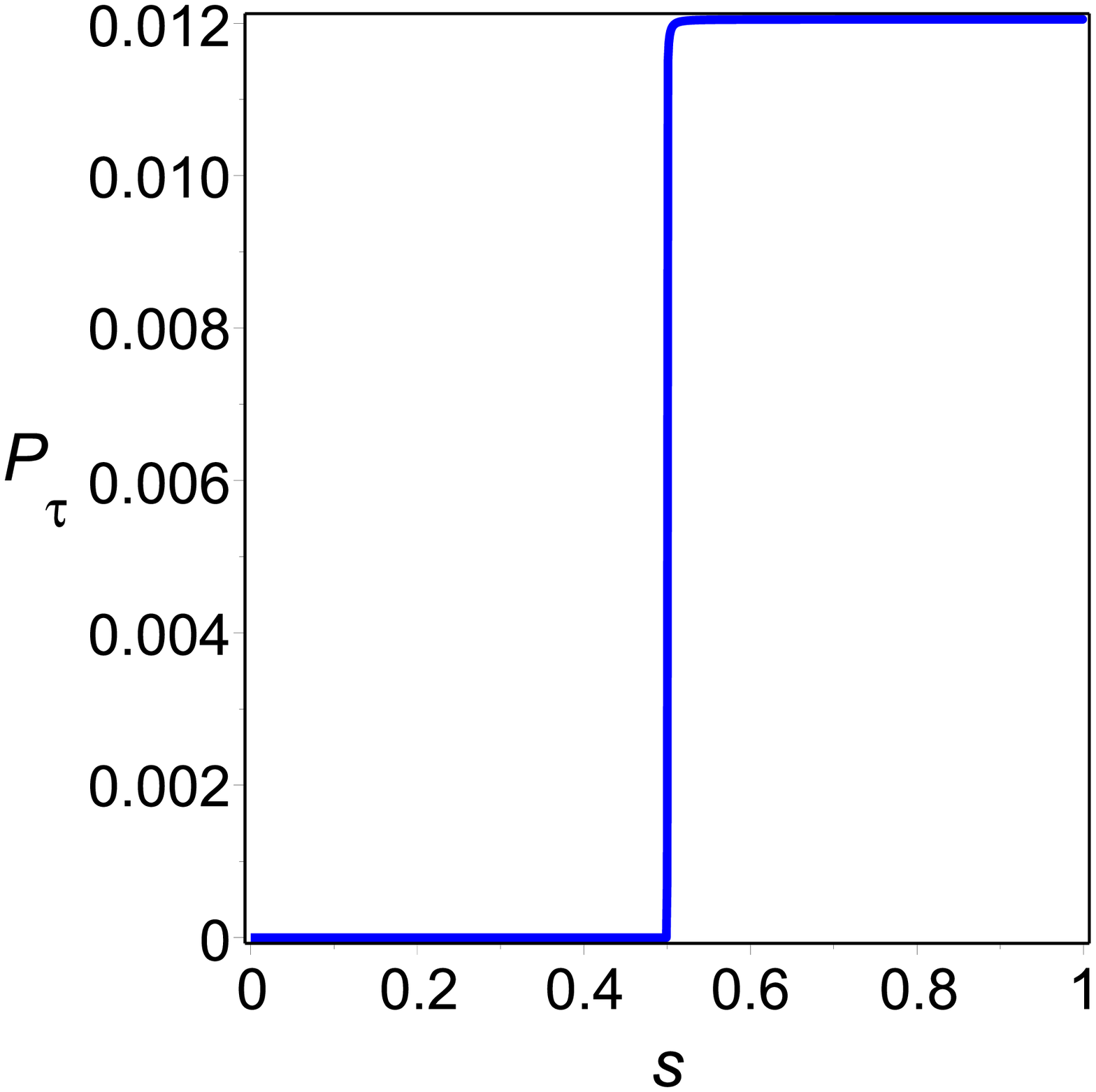}}
\scalebox{0.182}{\includegraphics{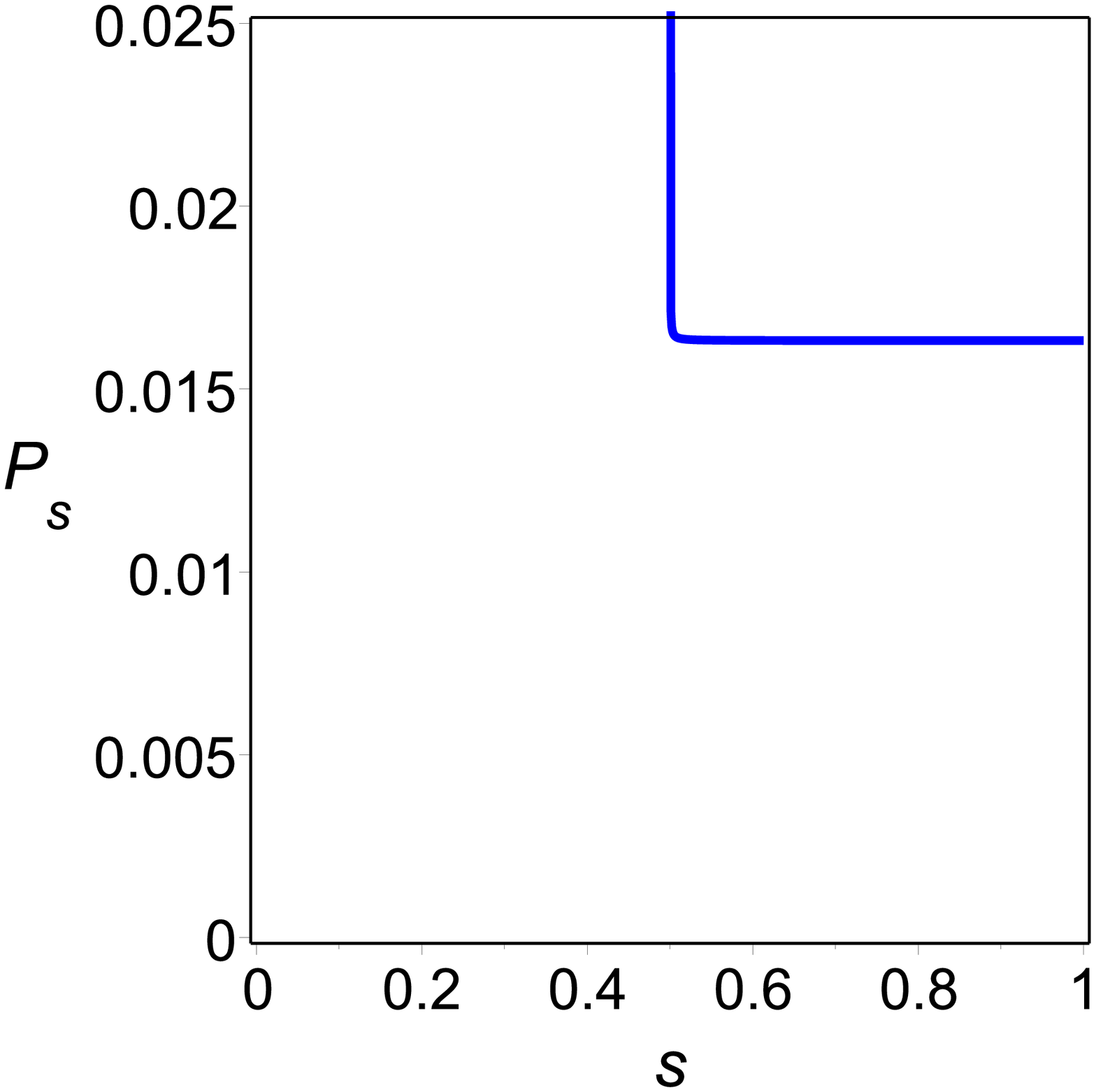}}
\caption{Nonlinear NQA.The transition probability (left panel) and the survival probability (right panel) as functions of the scaled time, $s =t/\tau$ ($g=2, \delta = 7.5\cdot10^{-5}, \tau=  5.5\cdot 10^{4}, N = 2^{40}$).}
\label{P4a}
\end{figure}

We rewrite the total time-dependent non-Hermitian Hamiltonian as,
\begin{align}\label{Eq10}
 {\mathcal H}_\tau(t) =  {\mathcal H}_0 +  h_0(1- f(t)){\mathcal H}_1,
\end{align}
where $ h_0 =( g+i\delta)$, and $f(t)$ is a monotonic function of $t$. For concreteness, we choose $g=2$, and impose the following boundary conditions: $f(0)=0$ and $f(\tau) =1$, where $\tau$ denotes the computational time.

We choose $f(t)$ as a solution of,
\begin{align}\label{Eq12}
  \frac{df}{dt} = \frac{\beta \delta}{\tau}\Big(1 + \Big(\frac{ (1-2f)}{\delta}\Big)^2\Big),
\end{align}
where $\beta = \arctan ({1}/{\delta})$. Performing the integration, we find
\begin{align}\label{Eq13}
 t =  \frac{\tau}{2} +\frac{\tau}{2\beta}\arctan\frac{ (2f-1)}{\delta}.
\end{align}
By inverting this function we obtain,
\begin{align}\label{Eq14}
f(t)  = \frac{1}{2}+ \frac{\delta}{2}\tan\Big( \beta \Big( \frac{2 t}{\tau}- 1\Big)\Big).
\end{align}
From here it follows that $f(\tau)=1$, and the computation time is $\tau$.

In Figs. \ref{P4} and \ref{P4a} we present the results of numerical calculations for different choice of parameters, $\delta$ and $\tau$. Our results show that the nonlinear NQA can be realized with the transition probabilities, $P_\tau \approx 1.2\cdot 10^{-2}$ and $P_s\approx 1.6\cdot 10^{-2}$. The computation time,  $\tau \approx 5.5\cdot 10^{4} $, is better than the time of quantum search predicted by the Grover algorithm, $\tau =\sqrt{N}\approx  10^{6}$ (for $n = 40$).

{\em Conclusion.}-- The field of quantum adiabatic computation is well-established, and many useful results are discussed in the literature. One of the main problems of this approach is that the energy gap between the ground state to be found and the excited states is generally exponentially small. This requires exponentially large computation times. On the other hand, in the dissipative (non-Hermitian) regime, the energy gap is defined by the relaxation parameters, and may not be exponentially small. (See also \cite{Laser1, Laser2}.) In this case, the computational time can be significantly reduced. But then, another problem appears--the quantum computer has a finite probability to be destroyed (which happens anyway). One way to overcome this problem was discussed in [14,15], where both dissipation and external pumping in the locked laser system was used to model the Ising system in its stationary ground state.
But still many theoretical and experimental issues must be resolved in order to  build this type of ``Ising machine".

The results presented in our paper demonstrate that non-Hermitian quantum computations can be used for two purposes. One is to use non-Hermitian quantum algorithms together with the use of classical computer to reduce computation time. We are in the process of demonstrating this option for some classes of Ising models. Another purpose is to build a real ``non-Hermitian quantum computer" (NHQC) to solve specific problems rapidly. As was demonstrated in this paper, there will be a tradeoff between the probability of finding the desired outcome and the probability of survival of the computer.   As our results show, there are useful ways to improve the performance of the NHQC.

~~~~~\\

The work by G.P. Berman was carried out under the auspices of the National Nuclear Security Administration of the U.S. Department of Energy at Los Alamos National Laboratory under Contract
No. DE-AC52-06NA25396.  A.I. Nesterov acknowledges the support from the CONACyT, Grant No. 118930.

\end {document}